\documentclass{PoS}

\title{ The bulk transition of many-flavour QCD and the search for a UVFP at strong coupling}

\ShortTitle{A strong coupling UVFP in many-flavour QCD?}

\author{Albert Deuzeman\footnote{Present address: Institute for Theoretical Physics, Ch-3012 Bern, Switzerland} and \speaker{Elisabetta Pallante}\\
        Centre for Theoretical Physics, University of Groningen, 9747 AG, Netherlands\\
        E-mail: \email{a.deuzeman@rug.nl, e.pallante@rug.nl}}

\author{Maria Paola Lombardo\\
        INFN-Laboratori Nazionali di Frascati, I-00044, Frascati (RM), Italy\\
        E-mail: \email{mariapaola.lombardo@lnf.infn.it}}

\abstract{We explore the nature of the bulk transition observed at strong coupling in the $SU(3)$ gauge theory with $N_f=12$ fermions in the fundamental representation. The transition separates a weak coupling chirally symmetric phase from a strong coupling chirally broken phase and is compatible with the scenario where conformality is restored by increasing the flavour content of a non abelian gauge theory. We explore the intriguing possibility that the observed bulk transition is associated with the occurrence of an ultraviolet fixed point (UVFP) at strong coupling, where a new theory emerges in the continuum.}

\FullConference{The XXVIII International Symposium on Lattice Field Theory, Lattice2010\\
		June 14-19, 2010\\
		Villasimius, Italy}

\begin{document}

\section{Theoretical premise}

Conformal symmetry restoration in non abelian gauge theories opens new possibilities for physics beyond the Standard Model, allowing for mechanisms of electroweak symmetry breaking induced by strongly coupled theories at the TeV scale. These theories offer an alternative to weakly coupled supersymmetry. They possibly live in a quasi-conformal region of the parameter space, as it is true for walking technicolor, and are the direct descendants of SU(N) Yang-Mills with $N=3$ colours and $N_f$ Dirac fermions in the fundamental or adjoint representation.
In \cite{us:PRD} we have provided evidence for the restoration of conformality in $N_f=12$ flavour-QCD. A strategy, based on previous  theoretical work \cite{Miransky, Appelquist_CW}, was devised in order to indirectly probe the emergence of the non trivial conformal infrared fixed point (IRFP), possibly at strong coupling. 

\begin{figure}[b]
\begin{center}
\includegraphics[width=.4\textwidth]{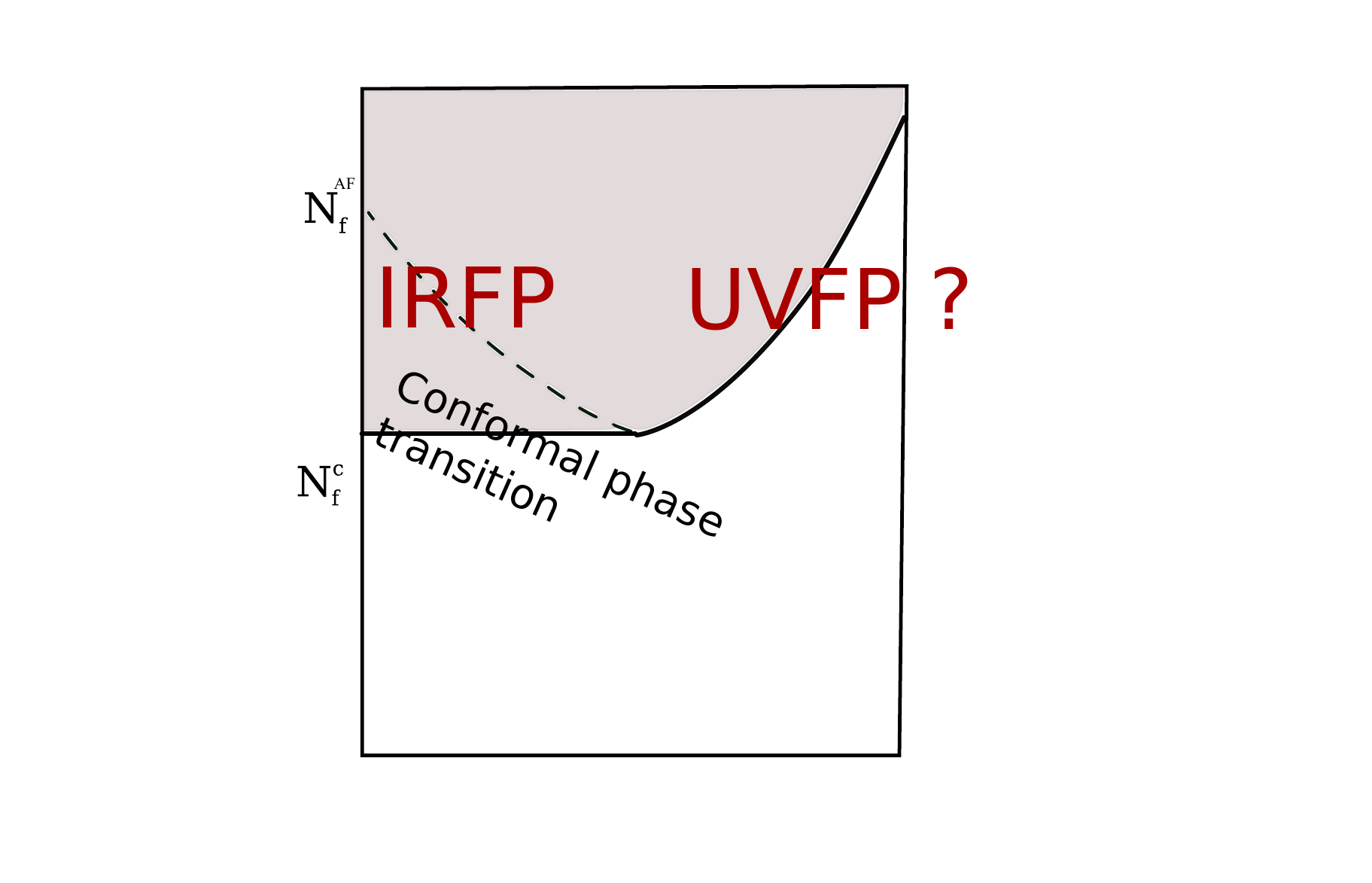}
\caption{The zero temperature $N_f\, -\, g$ phase diagram proposed in refs.~\cite{Miransky,Appelquist_CW} and supported by a lattice study~\cite{us:PRD}. The region $N_f^c<N_f<N_f^{AF}$ indicates the "conformal window". The (brown) shaded area is where chiral symmetry is restored. 
We investigate the nature of the bulk transition line at strong coupling and explore the possible existence of a UVFP, as conjectured in \cite{BZ} and recently revisited in \cite{Kaplan}.}
\label{fig:cw}
\end{center}
\end{figure}

Within this picture various key features emerge. Conformality can a priori be realized with or without the presence of a so called ``conformal window''. The first scenario was introduced in \cite{Appelquist_CW,Miransky}, while the second was earlier on 
 conjectured by Banks and Zaks \cite{BZ}. If a conformal window is present -- see Fig.~\ref{fig:cw} -- a family of theories with $N_f^c<N_f<N_f^{AF}$ exists, for which chiral symmetry is exact and the theory is deconfined on both sides of the IRFP line. The critical number of flavours $N_f^c$ is marked by a conformal phase transition and signals the opening 
of the conformal window. 
The lattice study in \cite{us:PRD} favours the existence of a conformal window, see also \cite{lat2010:MP} for an updated summary.

Another interesting aspect emerges when moving towards stronger coupling. 
For both scenarios, the lattice theory is expected to undergo a transition to a chirally broken phase at sufficiently strong bare lattice coupling (indicated as $g$ or equivalently $\beta\equiv 6/g^2$ in the following). This is referred to as a bulk transition, a zero temperature transition which interests all scales in the system. 
If the bulk transition is of second order, with non trivial exponents,
then an interacting continuum theory can be defined in correspondence of the strongly coupled UVFP. It is then natural to ask what is the nature of the bulk transition observed on the lattice: Is it a lattice artifact naturally associated with the strong coupling regime of the lattice theory, or does it signal the emergence of a UVFP in the continuum theory?
Recently, a mechanism for the loss of conformality has been suggested \cite{Kaplan}, inspired by general features of AdS/CFT and different from already explored solutions of supersymmetric QCD. The conformal window would disappear due to the pair annihilation of an IR and an UV fixed point. The IR and UV branches are the location of conformal theories in $d=4$, dual to each other.
We add that the UVFP in the continuum theory could also emerge just at the very end-point of the conformal window. In this case, the lattice realization of the theory could show a strong coupling line of bulk transitions of first order or crossover, while its end-point will necessary be a critical point where a second order transition will occur. 
In section \ref{sec:bulk} we report on lattice simulations with $N_f=12$ fundamental flavours in the region of the bulk transition observed in \cite{us:PRD}. Here, we further explore the nature of the transition. We first scrutinise its temperature dependence, and we once more exclude its thermal nature in agreement with \cite{us:PRD}. In section \ref{sec:scaling} we present a preliminary study of the mass dependence and volume scaling of the chiral order parameter, its susceptibilities and the plaquette operator. We conclude with an interpretation of the results and point to open questions.

\section{The bulk transition with $N_f=12$ fundamental flavours}
\label{sec:bulk}
\begin{figure}[t]
\begin{center}
\includegraphics[width=.60\textwidth]{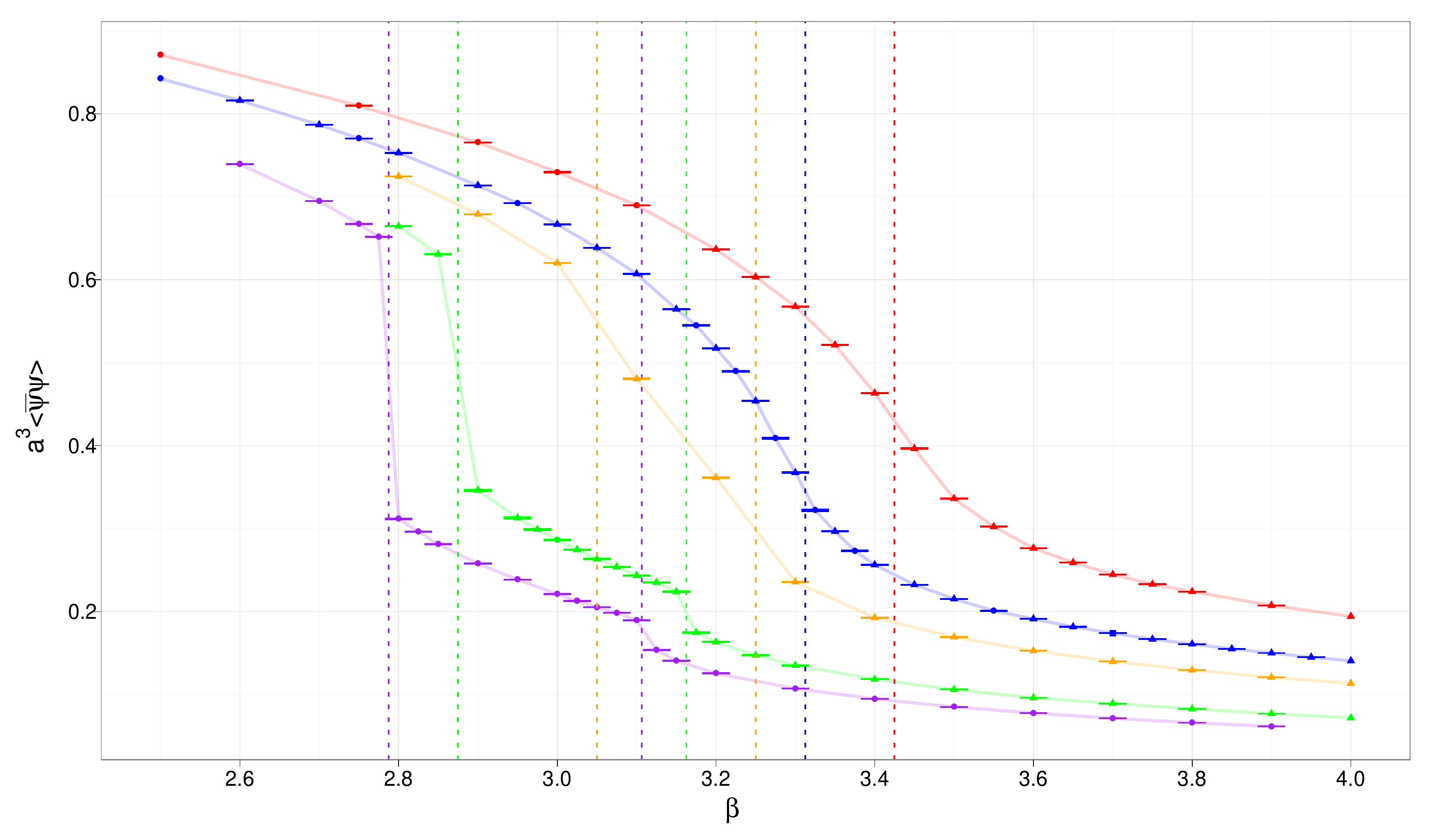}
\caption{  The chiral condensate from cold starts on $24^4$ volumes as a function of the lattice coupling $\beta$ for varying fermion bare mass, from left to right $am$=0.020 (purple), 0.025 (green), 0.040 (orange), 0.050 (blue) and 0.070 (red). Vertical lines are the positions of the peak of the finite difference approximation to the derivative of the condensate. Data at $am=0.020$ and 0.025 are preliminary and certainly not thermalized  in the region of the second, weak rapid crossover around $\beta =3.1$.  }
\label{fig:51}
\end{center}
\vspace{-0.4cm}
\end{figure}

\begin{figure}
\begin{minipage}[t]{0.5\linewidth}
\begin{center}
\includegraphics[scale=0.22]{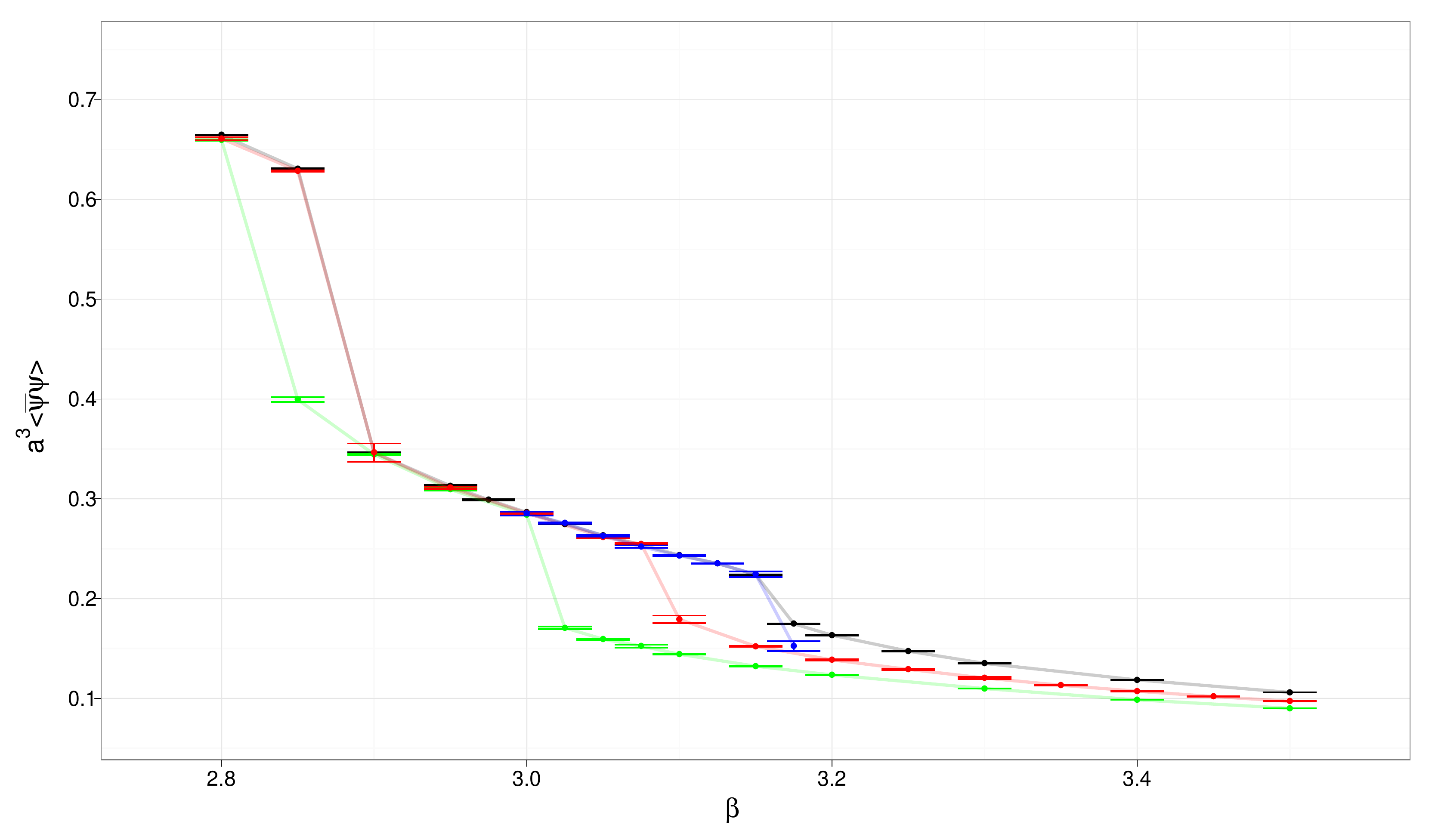} 
\caption{  The chiral condensate from cold starts at $am=0.025$, as a function of the temporal extent, from left to right $N_t =$ 6 (green), 8 (red), 10 (blue) and 24 (black). The results settle for $N_t\geq 10$. Same caveats as in Fig.~2 apply. }
\label{fig:52}
\end{center}
\end{minipage}
\hspace{0.2cm}
\begin{minipage}[t]{0.5\linewidth}
\begin{center}
\includegraphics[scale=0.22]{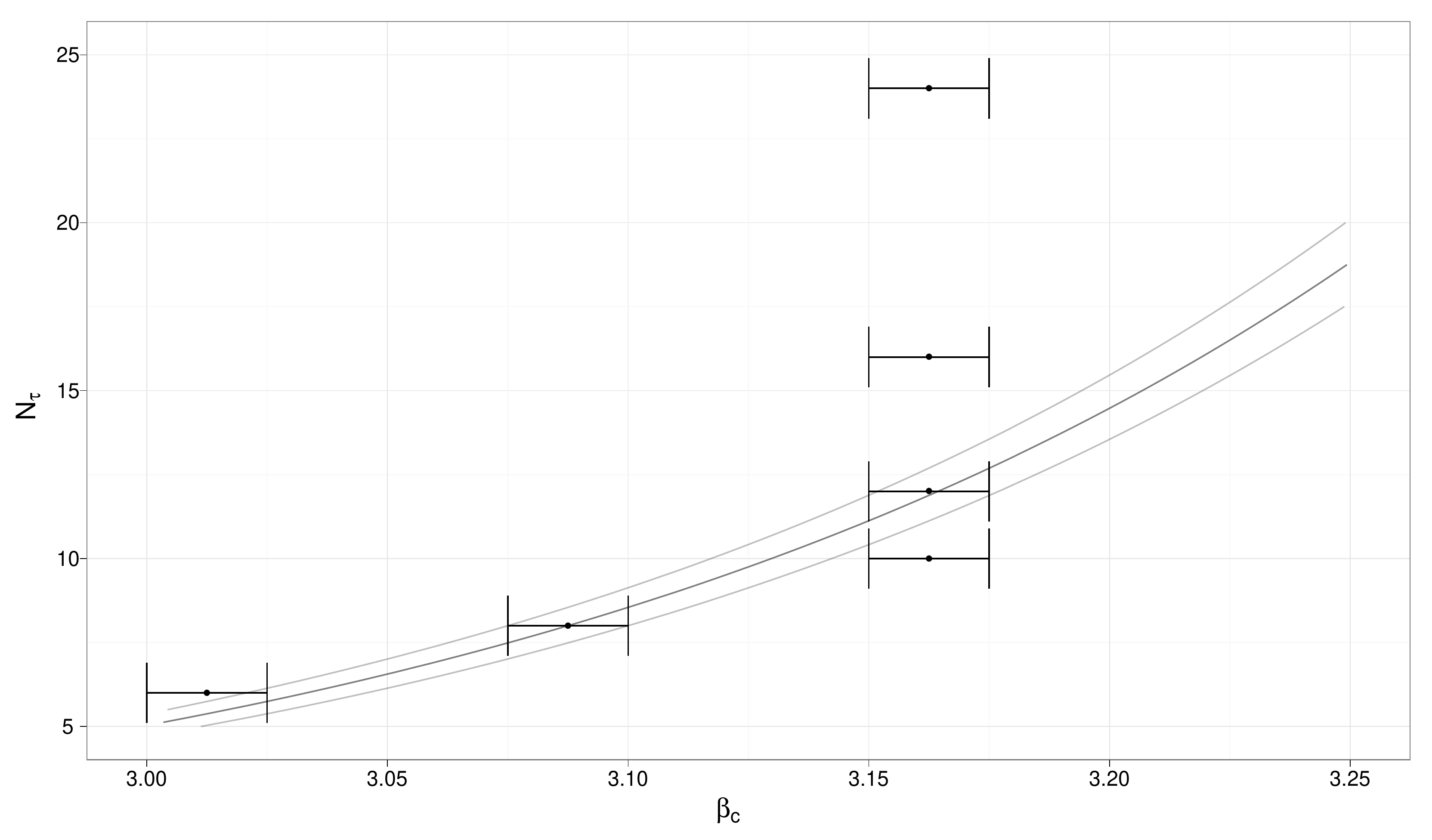}
\caption{$N_t -\beta_c$ plot determined from a cold start, for $am=0.025$ and $N_t = 6,8,10,12,16,24$. The results become insensitive to $N_t$ for $N_t\geq 10$. The agreement with the perturbative scaling prediction (superimposed curve) would point at a thermal nature of the transition, and this is not the case.}
\label{fig:53}
\end{center}
\end{minipage}
\end{figure}

A first look at the mass dependence of the candidate bulk transition highlights a few interesting properties. First, since the shielding effect of dynamical fermions increases by lowering their mass we expect the chiral symmetry breaking transition to move towards stronger couplings while approaching the chiral limit; this is indeed observed in Fig.~\ref{fig:51}. 
The transition clearly becomes stronger at  the lightest explored masses, $am \lesssim 0.04$. 
The preliminary data in Fig.~\ref{fig:51}, all collected from cold starts, show a pronounced 'jump' at stronger coupling and a tiny 'jump' at weaker coupling with strong instabilities in the whole region between them; when performing thermalization checks from hot starts we noticed lack of thermalization for several couplings. These qualitative features already suggest the presence of a first order transition, possibly accompanied by a large hysteresis loop, and the smaller jump at weaker coupling could be due to incomplete sampling of metastable states. 
A scaling study by varying the temperature of the system is in this case crucial in order to draw conclusions. We thus repeat what we did in \cite{us:PRD}, on an extended range of masses. Fig.~\ref{fig:52} shows that the location of both jumps settles for temporal extents $N_t\geq 10$, while Fig.~\ref{fig:53} shows that no perturbative scaling curve can fit the location of the weak coupling jump\footnote{The same behaviour is also observed for the strong coupling jump, thus also excluding its thermal nature. Notice also that no plausible scenario can justify its thermal nature, if a second transition at weaker coupling exists.}, thus excluding its thermal nature in agreement with \cite{us:PRD}.
\begin{figure}
\begin{center}
\includegraphics[width=.60\textwidth]{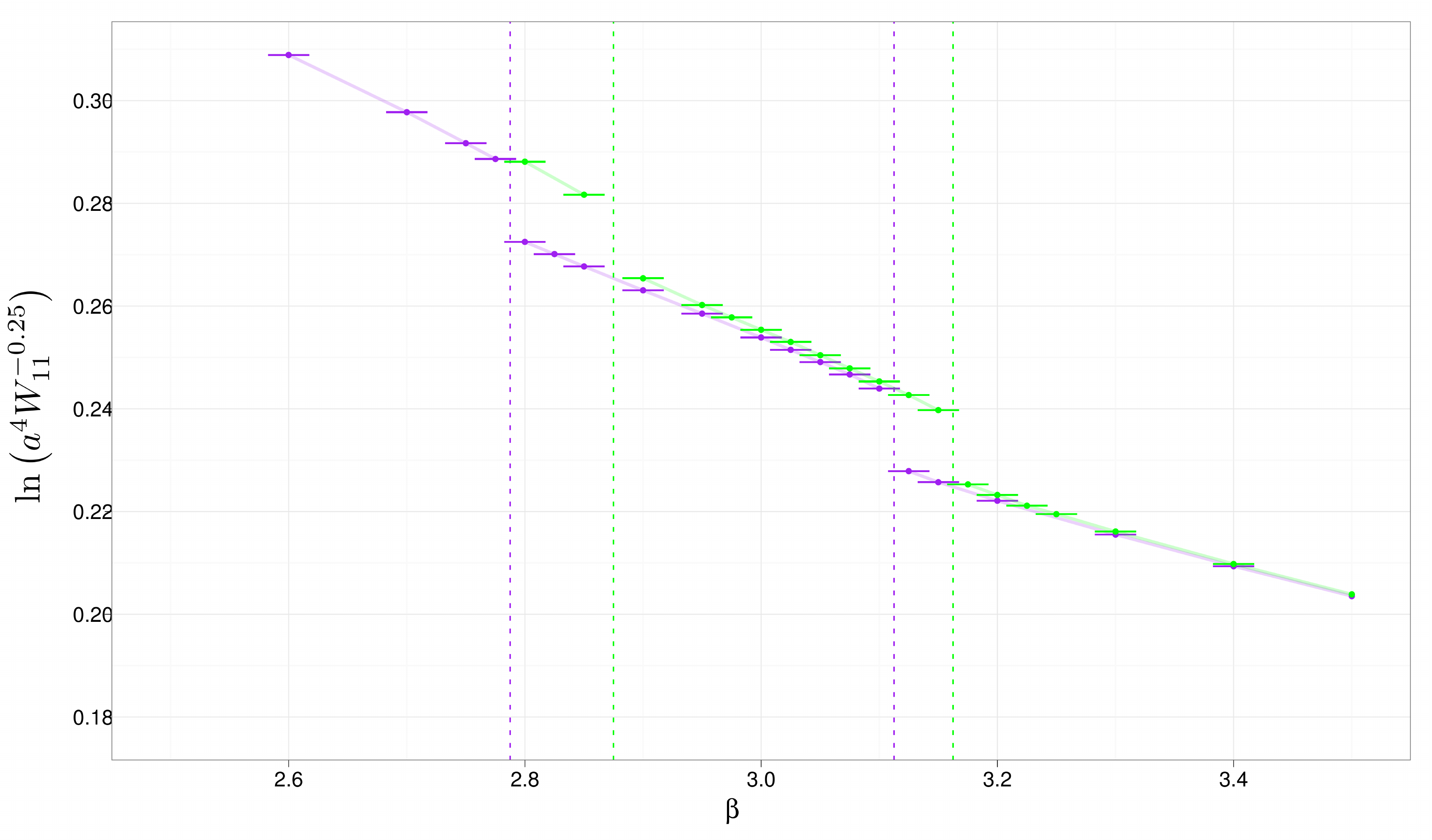}
\caption{A transform of the plaquette, also useful for perturbative studies, defined as $\ln (a^4 W_{11}^{-0.25})$ from the average of $a^4 W_{11}$ as a function of $\beta$. Data are from cold starts, at volumes $24^4$ and masses $am$ = 0.020 (purple, leftmost) and 0.025 (green, rightmost). Vertical lines indicate the regions or rapid crossover also observed for the chiral condensate. Strong instabilities are observed in the whole region between the two crossovers. Results are preliminary, see the text for a discussion.}
\label{fig:56}
\end{center}
\vspace{-0.4cm}
\end{figure}
The next task of our ongoing study is thus the one of establishing the location and nature of the bulk phase transition(s) -- one or two of them -- and their fate in the chiral limit, including a finite size scaling analysis. At this stage, given the preliminary nature of the data and the clear lack of thermalization of the light mass results around $\beta =3.1$, we cannot draw firm conclusions. However, in Section \ref{sec:scaling} we will show that the observed 'jump' at strong coupling and the behaviour of the cumulant and chiral susceptibilities are strongly suggestive of
a first order chiral transition. 
A more detailed study of the region between the strong coupling 'jump' and the weak coupling tiny 'jump' around $\beta =3.1$ at the lightest masses will be necessary to confirm this picture, and to clarify if the latter is simply due to the lack of thermalization, partial sampling of a large hysteresis loop, or maybe a genuinely separate structural lattice transition.
None of our data seem to support the occurrence of a second order phase transition in the chiral limit, and thus the emergence of a UVFP.
One could also think of another possibility that could give rise to two 'jumps' in the chiral condensate at our lightest masses. It is well known that in a crossover region different
observables can show a rapid crossover for different values of the coupling, which will converge to a common value at the true transition point.
Would our two 'jumps' be confirmed, we would instead have signals of
comparable intensity for different observables -- chiral condensate and plaquette (see Fig.~\ref{fig:56}) -- and increasingly significant for decreasing masses, which might be suggestive of two distinct transitions also in the chiral limit. Again, which scenario is realized is the subject of our ongoing study.

In the following, we report on preliminary results for the Monte Carlo history of the chiral condensate, its susceptibilities, the chiral cumulant, and we collect evidences for the occurrence of a first order transition at the lightest masses and therefore in the chiral limit. 
\begin{figure}
\begin{minipage}{0.5\linewidth}
\begin{center}
\includegraphics[scale=0.22]{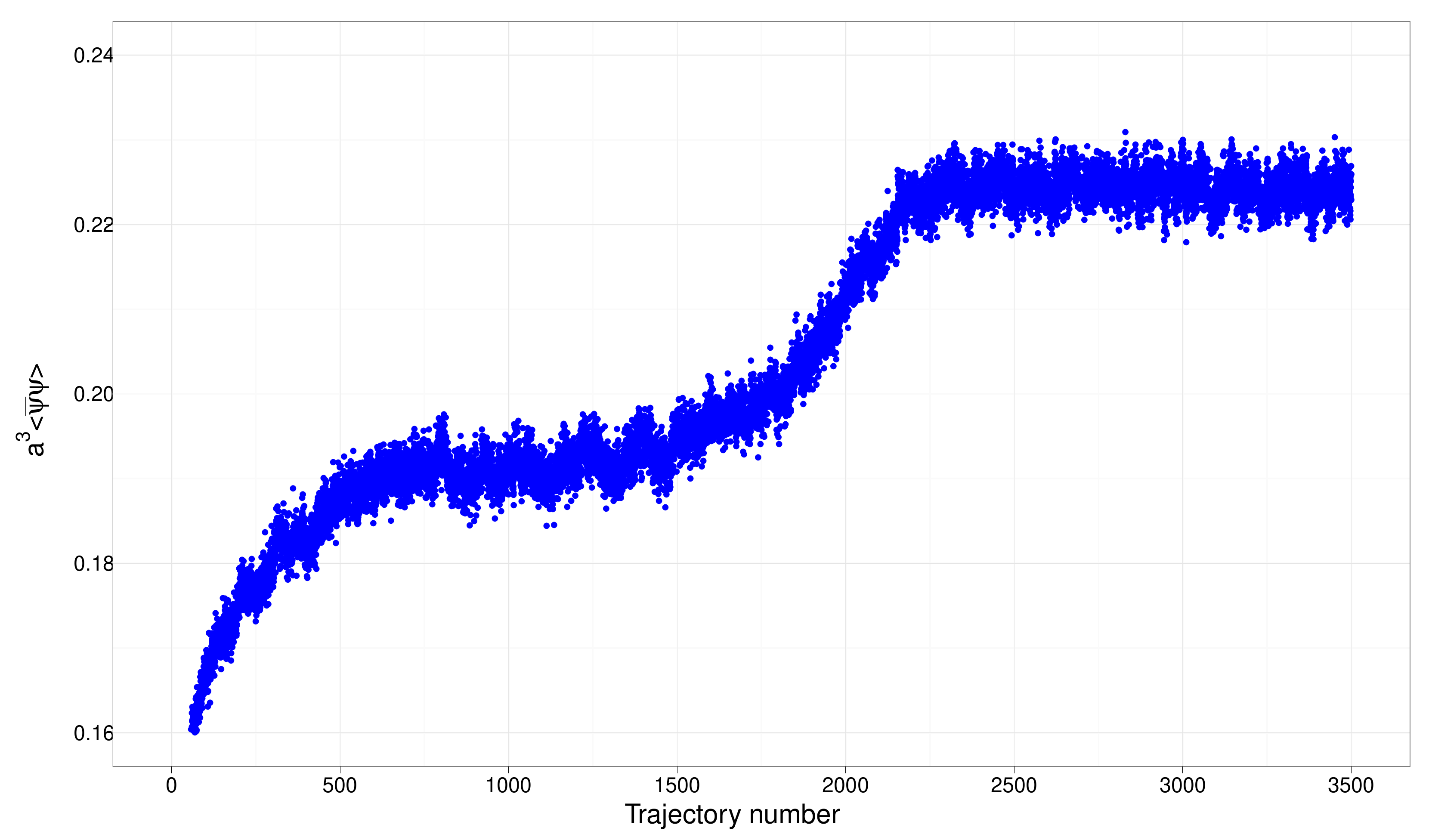}
\caption{Tunnelling of the chiral condensate observed for a cold start at $am =0.025$, volume $24^4$ and $\beta = 3.150$.  }
\label{fig:58}
\end{center}
\end{minipage}
\hspace{0.2cm}
\begin{minipage}{0.5\linewidth}
\begin{center}
\includegraphics[scale=0.22]{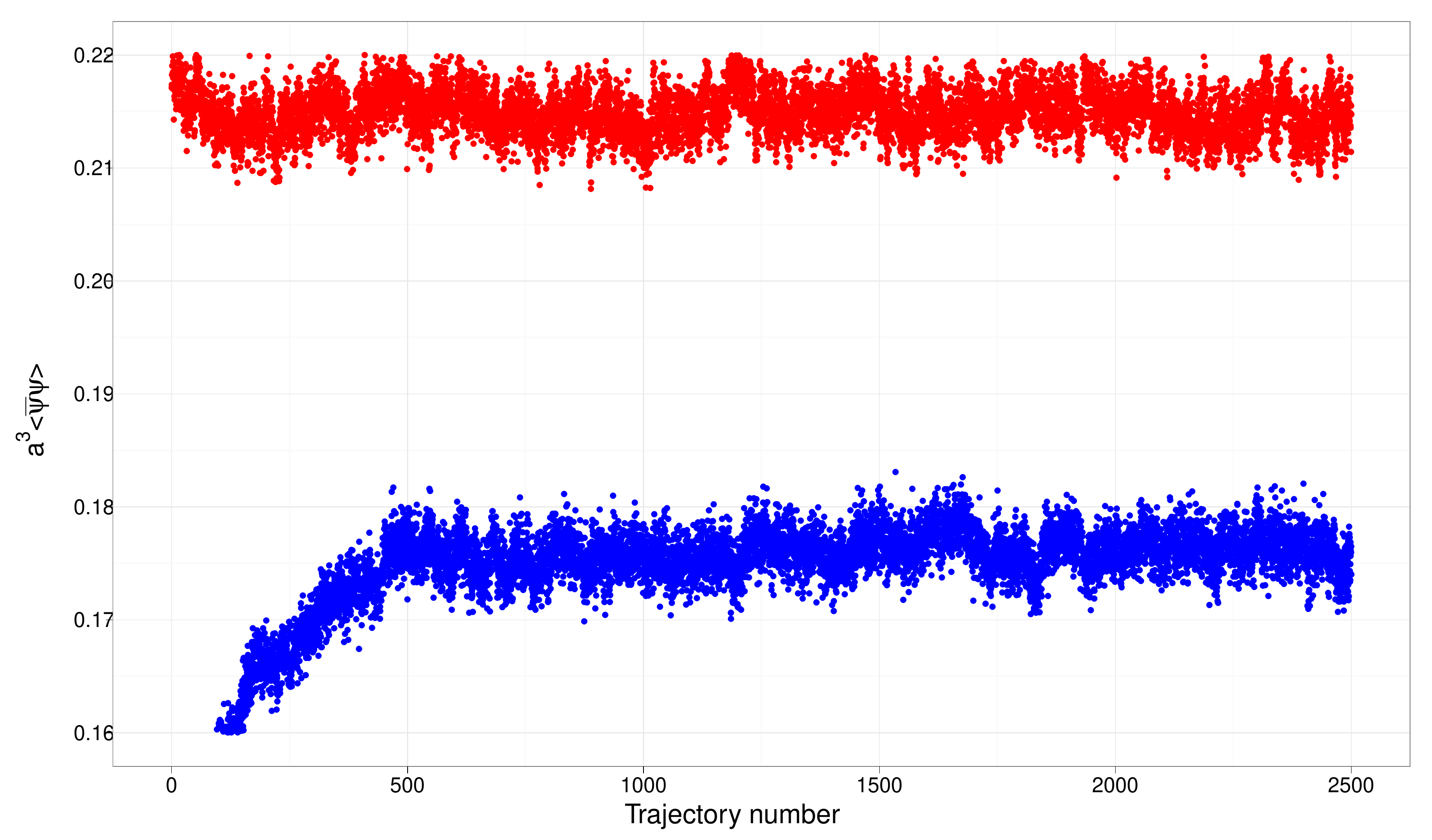}
\caption{Monte Carlo history of the chiral condensate at $\beta =3.175$, $am=0.025$ and volume $24^4$, from a cold start (blue, lower band) and a hot start (red, upper band).  The separation is maintained over at least 2000 trajectories. }
\label{fig:57}
\end{center}
\end{minipage}
\end{figure}
\subsection{Hints for a first order chiral transition}
\label{sec:scaling}
\begin{figure}
\begin{minipage}{0.5\linewidth}
\begin{center}
\includegraphics[scale=0.22]{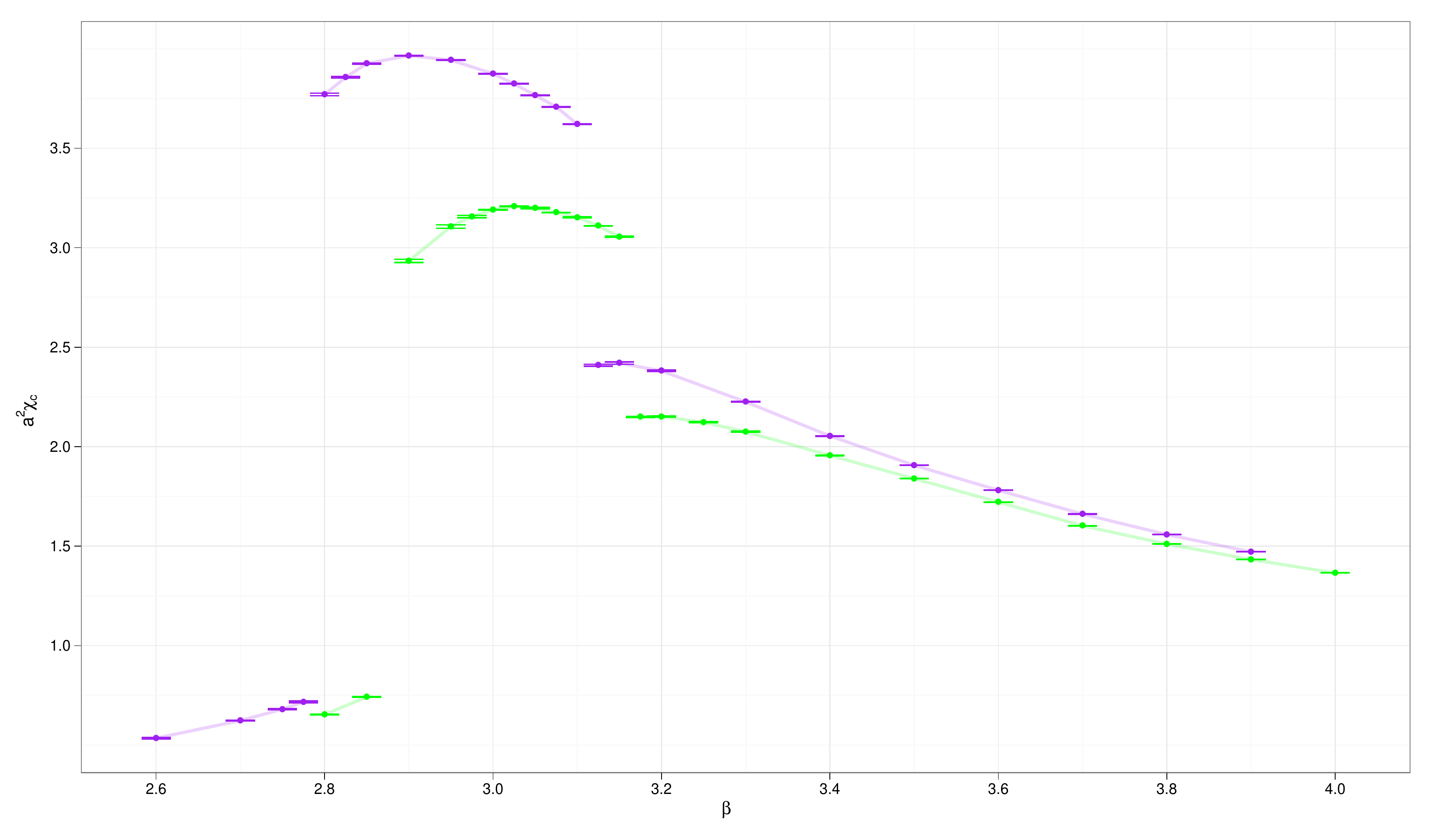}
\caption{The connected susceptibility from cold starts, at volumes $16^3\times 24$ and $24^4$, for  $am = 0.020$ (purple, upper) and 0.025 (green, lower). Quasi-discontinuities are in correspondence of the location of the chiral condensate (and plaquette) jumps, and need to be confirmed.}
\label{fig:59}
\end{center}
\end{minipage}
\hspace{0.2cm}
\begin{minipage}{0.5\linewidth}
\begin{center}
\includegraphics[scale=0.23]{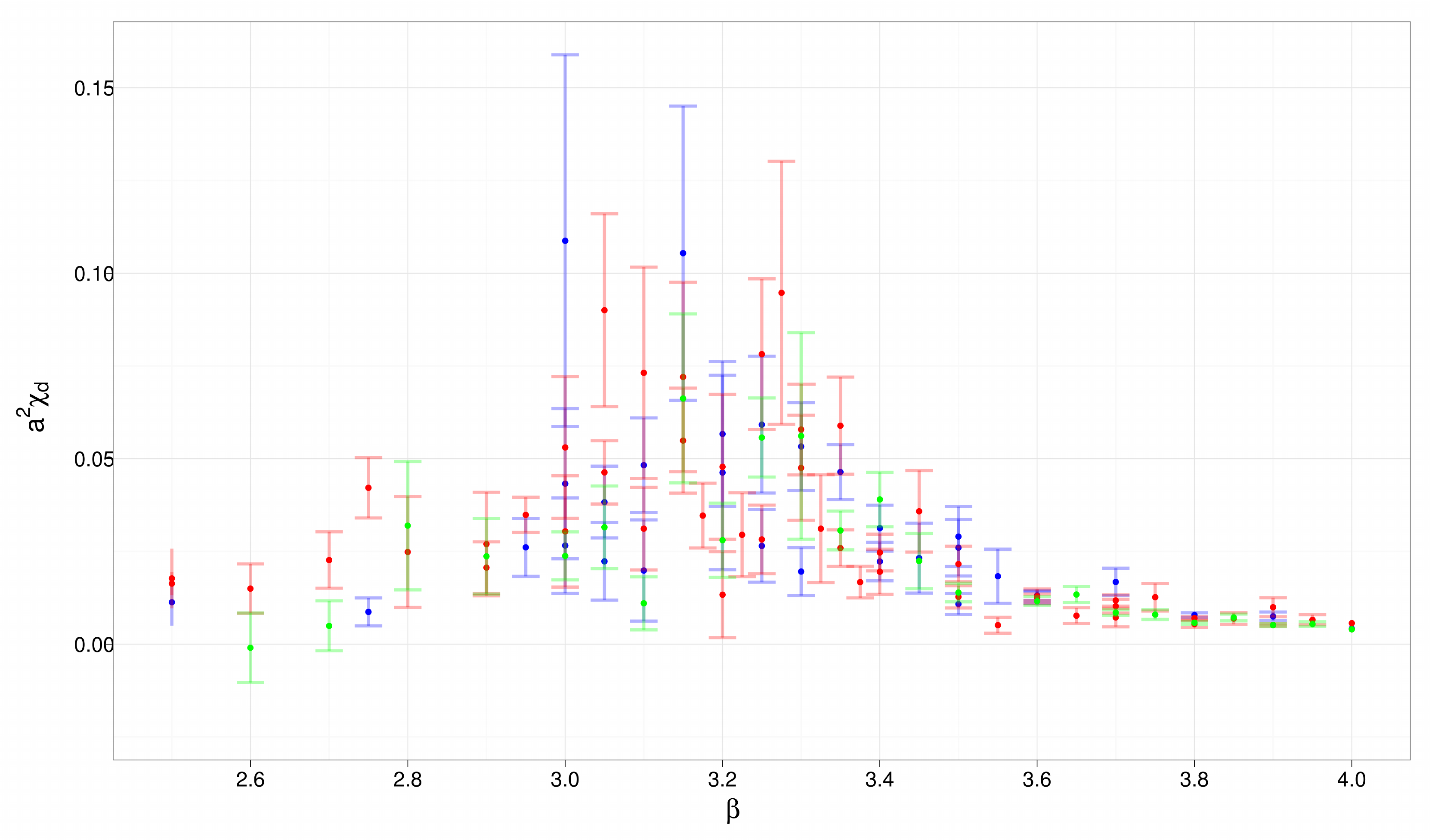}
\caption{The disconnected susceptibility at volumes $12^3\times 24$ (red), $16^3\times 24$ (green) and $24^4$ (blue), for  a mass $am = 0.05$.}
\label{fig:510}
\end{center} 
\end{minipage}
\end{figure}

The simulations setup for the present analysis has been described in \cite{us:PRD}. 
In Fig.~\ref{fig:57} we show evidence of a persistent separation between a cold and a hot start in the Monte Carlo history of the chiral condensate at one light mass $am=0.025$ and large volume $24^4$. A tunnelling between two values of the chiral condensate is observed in Fig.~\ref{fig:58}. 
The connected susceptibility from cold starts, in Fig.~\ref{fig:59}, is consistent with the rest of the observations. The apparent discontinuities might be associated with first order transitions, when not enough tunnelling between metastable states has occurred and not all branches of the hysteresis loop have been observed. It remains to be clarified what is the extent of the hysteresis region. 
\begin{figure}
\begin{center}
\includegraphics[width=.60\textwidth]{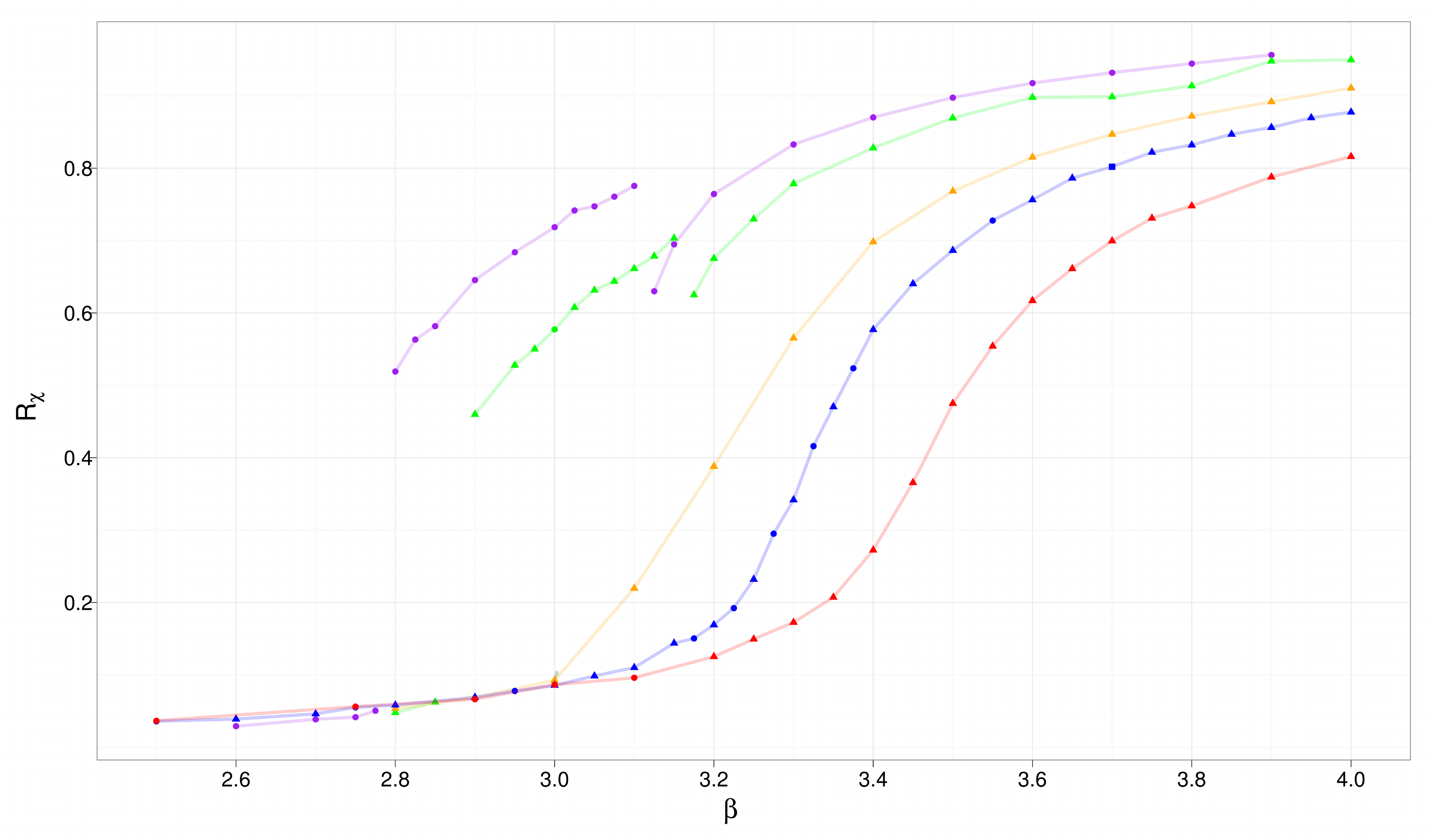}
\caption{The chiral cumulant $R_\chi$ at volumes $16^4$ from cold starts, for varying masses (from left to right) $am$ = 0.020 (purple), 0.025 (green), 0.04 (orange), 0.05 (blue) and 0.07 (red). Cumulants for the higher masses cross at a common value $\beta = 3.00(1)$.}
\label{fig:55}
\end{center}
\end{figure}
The chiral cumulant, defined as the ratio of the longitudinal and transverse susceptibilities $R_\chi = \chi_\sigma /\chi_\pi =(\chi_{conn}+\chi_{disc})/(\langle\bar\psi\psi\rangle /m)$, is a very useful quantity to explore the separation between a chirally broken and a symmetric phase.
The cumulants in Fig.~\ref{fig:55} approach one in the symmetric phase, as expected. For a second order phase transition the cumulant should be mass independent at the critical point $R_\chi (m,t=0) = 1/\delta$, with a value dictated by the critical exponent $\delta$, meaning that all cumulants will converge to the same value while approaching the critical point $t\to 0$. For a first order transition a similar behaviour can be expected with an exponent not constrained by universal scaling. Notice that the cumulant for the two lightest masses, and in the region between the weak and strong coupling jumps of the chiral condensate, clearly decreases for increasing mass values, as it is proper to a chirally symmetric phase. This suggests that chiral symmetry is only broken for $\beta \lesssim 2.8$.
At heavier masses, the transition is clearly weakened by mass effects. It might become a crossover, and in this case a second order phase transition at the end-point of the first order line should be observed. The disconnected susceptibility at a heavier mass $am=0.05$ in Fig.~\ref{fig:510} does not yet clearly show a volume dependent peak in the transition region; the statistics is simply too poor. 
\section{Conclusions and outlook}
While it is premature to draw definite conclusions, the present investigation of the bulk transition for the $SU(3)$ gauge theory with $N_f=12$ fundamental fermions favours a scenario where a strong first order transition separates a chirally symmetric phase at weaker coupling from a chirally broken phase at stronger coupling, at sufficiently light masses and thus in the chiral limit. The transition is weakened by heavier masses, as expected. It remains to be clarified if and for what mass values a crossover regime is entered. In this scenario, with a first order chiral transition, no continuum limit would exist. The occurrence of a chiral transition, independently of its nature, confirms the conclusion of \cite{lat2010:MP,us:PRD} that $N_f=12$ is in the conformal phase.
Further study at different volumes and enlarged range of light mass values is in progress; this aims at a proper finite size scaling analysis of the susceptibilities in the candidate critical region, the identification of the hysteresis loop(s), and establishing where a crossover regime is eventually entered at heavier masses.
Notice that if a first order nature of the transition is confirmed for $N_f=12$, this would not a priori exclude the occurrence of a UVFP before the closing of the conformal window; in fact, 
a UVFP could emerge just at the end-point of the conformal window. 
This work does not address the question:  would an additional four-fermion interaction change the picture? and how? One difference between the genuinely non-perturbative lattice formulation and the continuum lagrangian formulation of the theory is certainly that effective four-fermion operators will always be generated by gluon exchange at sufficiently strong coupling, but with a given functional dependence on the gauge coupling constant. Adding by hand a four-fermion term to the lattice action introduces one additional tunable coupling. Will a UVFP be found instead in this enlarged parameter space? Some intriguing theoretical questions also emerge: is the disappearance of conformality by merging of fixed points being realized in QCD or supersymmetric QCD? how does the backreaction of flavoured $4d$ branes affect possible mechanisms of disappearance of conformality in AdS/CFT?  And if a UVFP emerges, we should find its theory -- the dual of many-flavour QCD -- and how it differs from its supersymmetric cousin.

\acknowledgments 

This work was in part based on the MILC collaboration's public lattice gauge theory code. 
Simulations were performed on the Astron/RUG IBM BG/P in Groningen, and 
the IBM Power6+ Huygens at SARA (Amsterdam).


\begin{thebibliography}{99}



\bibitem{BZ}
  T.~Banks and A.~Zaks,
  Nucl.\ Phys.\  B {\bf 196} (1982) 189.
%
\bibitem{Miransky}
V.A.~Miransky, K.~Yamawaki, Phys.\ Rev.\ {\bf D55} (1997) 5051; Erratum-ibid.D56 (1997) 
3768. 
\bibitem{Appelquist_CW}
T. Appelquist, J. Terning, L.C.R. Wijewardhana, Phys. Rev. Lett. 77 (1996) 1214.

\bibitem{lat2010:MP}
A. Deuzeman, M.P. Lombardo, E. Pallante, PoS (Lattice 2010) 062.

\bibitem{us:PRD}
A. Deuzeman, M.P. Lombardo, E. Pallante, Phys. Rev. D82 (2010) 074503 [arXiv:0904.4662].

\bibitem{Kaplan}
 D.B. Kaplan, et al., Phys. Rev. D 80 (2009) 125005.


\end{thebibliography}
\end{document}